\documentstyle[aaspp4]{article}

\newcommand{\etal}{{et \it al.}}
\begin{document}

\title{ HST - NICMOS Color Transformations and Photometric Calibrations \altaffilmark{1} }

\author{Andrew W. Stephens\altaffilmark{2} and Jay A. Frogel\altaffilmark{2}}
\affil{Department of Astronomy, The Ohio State University, Columbus, Ohio 43210}

\author{Sergio Ortolani}
\affil{Universit\`a di Padova}

\author{Roger Davies}
\affil{University of Durham}

\author{Pascale Jablonka}
\affil{Observatoire de Paris-Meudon}

\author{Alvio Renzini}
\affil{ESO}

\and

\author{R. Michael Rich}
\affil{UCLA}

\altaffiltext{1}{Based on observations with the NASA/ESA Hubble Space Telescope obtained at the Space Telescope Science Institute, which is operated by AURA for NASA under contract NAS5-26555.}

\altaffiltext{2}{email: \{stephens,frogel\}@astronomy.ohio-state.edu}

\begin{abstract}

This paper presents color transformations for HST NICMOS camera 2 observations to the ground-based CIT/CTIO photometric system, 
using observations of nineteen moderately bright, red stars in Baade's window in the color range $0.7 < (J-K) < 1.6$.
We estimate an extension of the transformation to $0.4 < (J-K) < 3.0$ with five standards observed by STScI.   
Convolving near-IR spectra taken above the atmosphere with different filter transmission profiles, we simulate both NICMOS and ground-based photometry, obtaining results which are consistent with our transformation and its extension.

\end{abstract}


\section{Introduction}

The transformation of photometric observations from one filter system to another is rarely a trivial task.  This is especially true for observations of cool giants at near-IR wavelengths because of the presence of very deep molecular absorption bands both in the stars themselves (eg. CO and H$_2$O) and in the earth's atmosphere (eg. CO$_2$ and H$_2$O).  Examples of such transformations may be found in 
Bessell \& Brett (1988) and 
Elias \etal\ (1983).  Because of these molecular bands the transformation equations can be dependent not only on stellar color, but also on absolute luminosity and stellar metallicity.  For the Hubble Space Telescope (HST) Near Infrared Camera and Multi-Object Spectrograph (NICMOS) filters on camera 2 (NIC2; \cite{Mac97}) that have been used as pseudo-$JHK$ filters, the transformation to a standard ground-based system presents a particularly difficult case.  This is because the commonly used NIC2 filters (F110W for $J$, F160W for $H$, and F222M for $K$) differ significantly from the commonly used ground-based filters (\cite{Bes88}); also there are no telluric absorption features in HST observations.  Nonetheless, many scientific programs, including our own, will require accurate (uncertainties no worse than a few percent) knowledge of these transformations.

In order to derive accurate color transformations, we devoted two orbits of one of our observing programs (GO-7826) to gathering observations of cool, metal rich giants that have extensive ground-based observations on the CIT/CTIO system.  These giants are all in the Baade's Window field of the Galactic Bulge.  The brighter ones and a few of the fainter ones had previously been observed with the same single channel photometric system that was used to establish the CIT/CTIO grid of southern standards (\cite{Eli82}; \cite{Fro87}).  The remainder of the fainter stars had previously been observed with a NICMOS array by 
Tiede \etal\ (1995).

This paper presents our HST-NICMOS observations of these previously observed bulge giants which span the color range of $0.7 < (J-K) < 1.6$.  Section two outlines our observations; the reduction procedures are discussed in section three.  A comparison with ground-based measurements and the color transformations are in section four.  The final section is a brief summary.

\section{Observations}

The targets were originally observed on 1998, August 19 with HST, but problems encountered while traversing the South Atlantic Anomaly caused a loss of tracking.  Thanks to the gracious reallotment of time by the Telescope Time Review Board (TTRB), our fields were reobserved on 1998, October 28.  Of the first observation attempt, we were only able to confidently salvage the observations of the first target, thus our quoted BW1 measurements are actually the average of measurements made on both visits.

Our observations consist of eight pointings in Baade's Window (henceforth referred to as BW1-BW8), from which we have obtained measurements of 19 stars previously observed from the ground.  We will refer to these stars as ``standard'' stars.  The first field (BW1) is a small part of the ``BW4b'' field observed by \cite{Tie95} (see Fig. 1).  This field is fairly rich, and includes 12 stars observed by 
Tiede \etal\ (1995).  The remaining 7 fields (BW2-BW8) are of single stars observed previously by 
Frogel \& Whitford (1987).  The coordinates of each field are given in Table 1.

\begin{figure}
\plotone{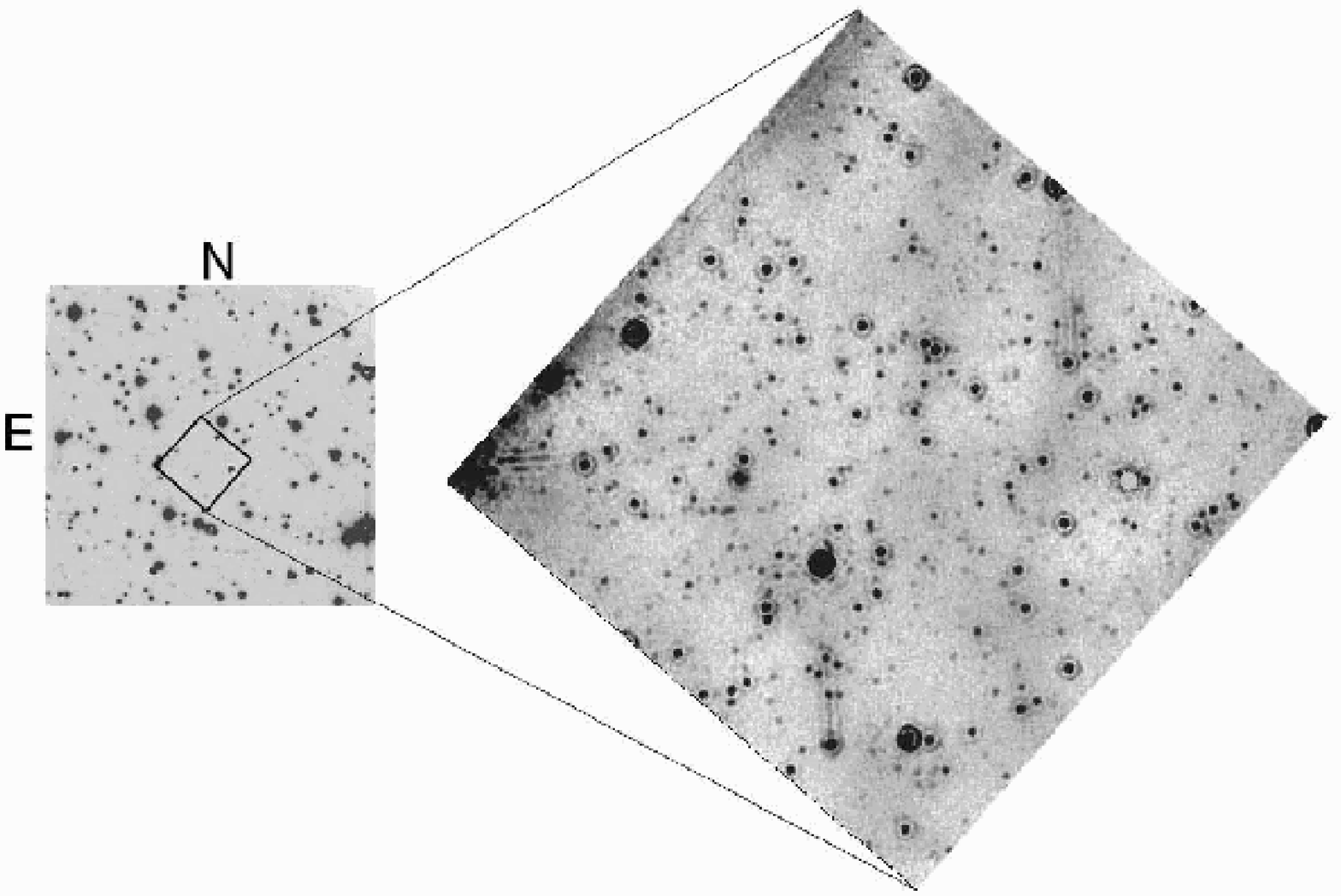}
\figcaption{The $K$-band BW1 field (right) in relation to the BW4b field of \cite{Tie95} (left).  The object in the west corner of the overlap region on the BW4b field is {\em not} a star, but an artifact of the reduction.  The ``hole'' in the west corner of the BW1 field is the NIC2 coronagraphic mask, which glows in the $K$-band and was therefore masked out during reduction.  The HST image has been stretched to bring out the fainter stars which results in showing the unevenness in the background.
\label{Fig_01}}
\end{figure}

\placefigure{Fig_01}

Our observations were taken with the NICMOS camera 2 (NIC2) which has a plate scale of $\sim$ 0\farcs 0757 pixel$^{-1}$ giving NIC2 a field of view of 19\farcs 4 on a side (376 arcsec$^2$).  The NICMOS focus was set at the compromise position 1-2, which optimizes the focus for simultaneous observations with cameras 1 and 2.

Each field was observed in three filters:  F110W ($\sim J$), F160W ($\sim H$), and F222M ($\sim K$), using a spiral dither pattern with 4 positions (see section 4 for a discussion of the filters).  We used 0\farcs 4 steps on BW1 to maximize the size of the overlapping field, and 5 \farcs0 steps on BW2-BW8 to minimize the effects of residual images from the bright stars.  

All of our observations used the {\sc multiaccum} mode (\cite{Mac97}) because of its optimization of the detector's dynamic range and cosmic ray rejection.  The BW1 field used the predefined sample sequences {\sc step2}, {\sc step8} and {\sc step16} with 12, 11, and 14 samples in $J$, $H$, and $K$ respectively, yielding exposure times of 18, 48, and 128 seconds.  Fields BW2-BW8 implemented the {\sc multiaccum} sample sequence {\sc scamrr}, designed for fast temporal sampling with a single camera, which performs a nondestructive read every 0.203 seconds, yielding typical exposure times of 0.41 seconds for each of the four dither positions.  Table 1 lists the total exposure times in each filter for all targets.

Cosmic ray (CR) rejection using the {\sc multiaccum} mode requires many intermediate reads.  Since most of our targets are of bright stars, our required exposure times were very short with only two to four intermediate reads during each exposure, an inadequate number for effective CR rejection.  We therefore rely on the four dither positions for CR and bad-pixel rejection.

\begin{deluxetable}{c c c r r r} 
\tablecaption{Observations}
\footnotesize
\tablewidth{0pt}
\tablehead{
	\colhead{Field}		&
	\colhead{$\alpha$}	&
	\colhead{$\delta$}	&
	\multispan3{Total Integration Time (s)} \\
	\colhead{} 		&  
	\colhead{(2000)}	&
	\colhead{(2000)}	&
	\colhead{F110W} 	&
	\colhead{F160W} 	&
	\colhead{F222M} 
}
\startdata
BW1 &  18 03 49.9 & -30 02 06.6 & 71.76 & 191.80 & 511.68 \\
BW2 &  18 03 52.6 & -30 01 23.1 &  1.64 &   1.64 &   1.64 \\
BW3 &  18 03 46.1 & -29 59 12.5 &  1.64 &   1.64 &   1.64 \\
BW4 &  18 03 34.1 & -29 59 58.5 &  2.44 &   1.64 &   1.64 \\
BW5 &  18 03 46.1 & -30 02 23.6 &  1.64 &   1.64 &   1.64 \\
BW6 &  18 03 44.3 & -30 03 41.7 &  3.24 &   1.64 &   1.64 \\
BW7 &  18 03 43.7 & -30 05 15.8 &  3.24 &   1.64 &   1.64 \\
BW8 &  18 03 33.3 & -30 05 21.4 &  1.64 &   1.64 &   1.64 \\
\enddata
\normalsize
\end{deluxetable}

\section{Reduction}

Our data were reduced using three different techniques.  The first technique was the standard STScI pipeline.  Afterwards we performed our own reductions with the NICRED v. 1.8 package dated 01/29/99 (\cite{Mcl97}, \cite{Leh99}), and with the STScI pipeline using the IRAF NICPROTO package (May 1999) to eliminate any residual pedestal.  All the reductions use flat fields and dark frames provided by STScI, although we generated our own bad pixel masks for the NICRED and NICPROTO reductions.  Not too surprisingly, all of these methods give similar photometric results, with no apparent systematic deviations and typical dispersions of $\sim 0.05$ magnitudes.  Although the images produced by NICRED appear cleaner in terms of CRs and bad pixels, all of our photometric measurements have been made off the frames reduced with the STScI pipeline and IRAF NICPROTO package.

Aperture photometry was performed according to the guidelines set down by STScI 
\verb8(http://www.stsci.edu/ftp/instrument_news/NICMOS/nicmos_doc_phot.html)8 
for point-sources observed with NIC2.  In our case we used the IRAF PHOT routine with a 0\farcs 5 aperture and a $1''$ sky annulus placed directly outside the aperture, assuming a NIC2 plate scale of 0\farcs0757 pixel$^{-1}$.  The measured count rate of each star was multiplied by 1.15 to correct to an infinite aperture, and then converted to flux using the NICMOS photometric keywords released December 1, 1998, obtained from \verb8http://www.stsci.edu/ftp/instrument_news/NICMOS/NICMOS_phot/keywords.html8, and listed in Table 2.  Finally the fluxes were converted to magnitudes using the parameters shown in Table 2.  (Note that the PHOTZPTs in Table 2 are {\em not} those given by STScI, but have been calculated to force the transformation to the CIT/CTIO system to be zero at zero ($m_{110}-m_{222}$) color, as can be seen in eqn. 6.)  One should also note that there are several methods of converting DN to magnitudes.  Two ways are to use either PHOTFLAM \& PHOTZPT or PHOTFNU \& ZP(VEGA), which are not equivalent and give different magnitudes (because of a zero-point shift).  We present our NICMOS magnitudes (ie. $m_{110}$, $m_{160}$, $m_{222}$) as obtained using PHOTFNU \& ZP(VEGA) so that we can directly compare our observations with the STScI standards (see Table 4), but we have calculated the transformations for both methods (eqns. 4-7).  Another method, and the easiest of these presented, is to use our transformations to go directly from instrumental magnitudes to CIT/CTIO magnitudes (see eqns. 8-9).

\begin{deluxetable}{c c c c c}
\tablecaption{NICMOS Camera 2 Photometry Parameters}
\footnotesize
\tablewidth{0pt}
\tablehead{
	\colhead{Filter}	&
	\colhead{PHOTFLAM}	&
	\colhead{PHOTZPT}	&
	\colhead{PHOTFNU}	&
	\colhead{ZP(VEGA)}	\\
	\colhead{}		&
	\colhead{(erg cm$^{-2}$ \AA$^{-1}$ DN$^{-1}$)} &
	\colhead{(mag)}		&
	\colhead{(Jy s DN$^{-1}$)} &
	\colhead{(Jy)}		
}
\startdata
F110W  & 4.292364E-19	& 23.566  & 1.823290E-6	& 1898 \\
F160W  & 2.406112E-19 	& 24.546  & 2.070057E-6	& 1113 \\
F222M  & 3.217631E-19	& 26.131  & 5.280848E-6	&  653 \\
\enddata
\normalsize
\end{deluxetable}

\section{Comparison with Ground-Based Observations}

Our measurements of the faint stars in our BW1 field are compared with the observations of 
Tiede \etal\ (1995).  Their observations were made on the 2.5m DuPont telescope at Las Campanas Observatory with the IRCAM (\cite{Per92}), which used a $256 \times 256$ HgCdTe NICMOS 3 detector with a plate scale of 0.348 arcseconds pixel$^{-1}$.  Their measurements were calibrated from 11 stars measured previously with single-channel photometry (\cite{Fro87} -- the same system used for the ground-based measurements of the stars in BW2-BW8) to the CIT/CTIO system (\cite{Eli82}, 
1983).  These measurements have reported average errors given in Fig. 3 of \cite{Tie95} ($\sigma_K(16) \sim 0.1$), and calibration errors of $\sigma_J = 0.02$, $\sigma_H = 0.03$, $\sigma_K = 0.02$ magnitudes.

Our measurements of the single stars in fields BW2-BW8 are compared with the measurements of 
Frogel \& Whitford (1987).  These were obtained on the CTIO 4m and the D3 InSb system, and transformed to the CIT/CTIO standard system.  Note that their published Table 1 is corrected for reddening and extinction by $K_0=K-0.14$, $(J-K)_0=(J-K)-0.26$, and $(H-K)_0=(H-K)-0.09$, while our values are uncorrected.

In Figure 2 we have plotted the difference between the ground-based measurements and our NICMOS PHOTFNU \& ZP(VEGA) calibrated measurements against color for each filter.  These plots show that a color term is required to bring the NICMOS data into agreement with ground-based observations.  By applying a linear fit to the data, giving stars fainter than $m_K = 10$ half weight (points with $(m_{110}-m_{222})<1.6$), we have calculated the appropriate transformations (eqns. 4-9).   As discussed in section 3, STScI supplies two sets of keywords for converting count rates to magnitudes.  To eliminate confusion, equations 1-3 explicitly state how the magnitudes used in calculating our transformations were determined.  

\begin{eqnarray}
m({\rm RAW})      = & -2.5 \times \log ( {\rm CR}_{1/2} )					           \\
m({\rm PHOTFLAM}) = & -2.5 \times \log ( {\rm CR}_{1/2} \times 1.15 \times {\rm PHOTFLAM}) - {\rm PHOTZPT} \\
m({\rm PHOTFNU})  = & -2.5 \times \log ( {\rm CR}_{1/2} \times 1.15 \times {\rm PHOTFNU}   / {\rm ZP(VEGA)} )
\end{eqnarray}
where CR$_{1/2}$ is the count rate measured in a 0\farcs5 aperture.

\begin{figure}
\plotone{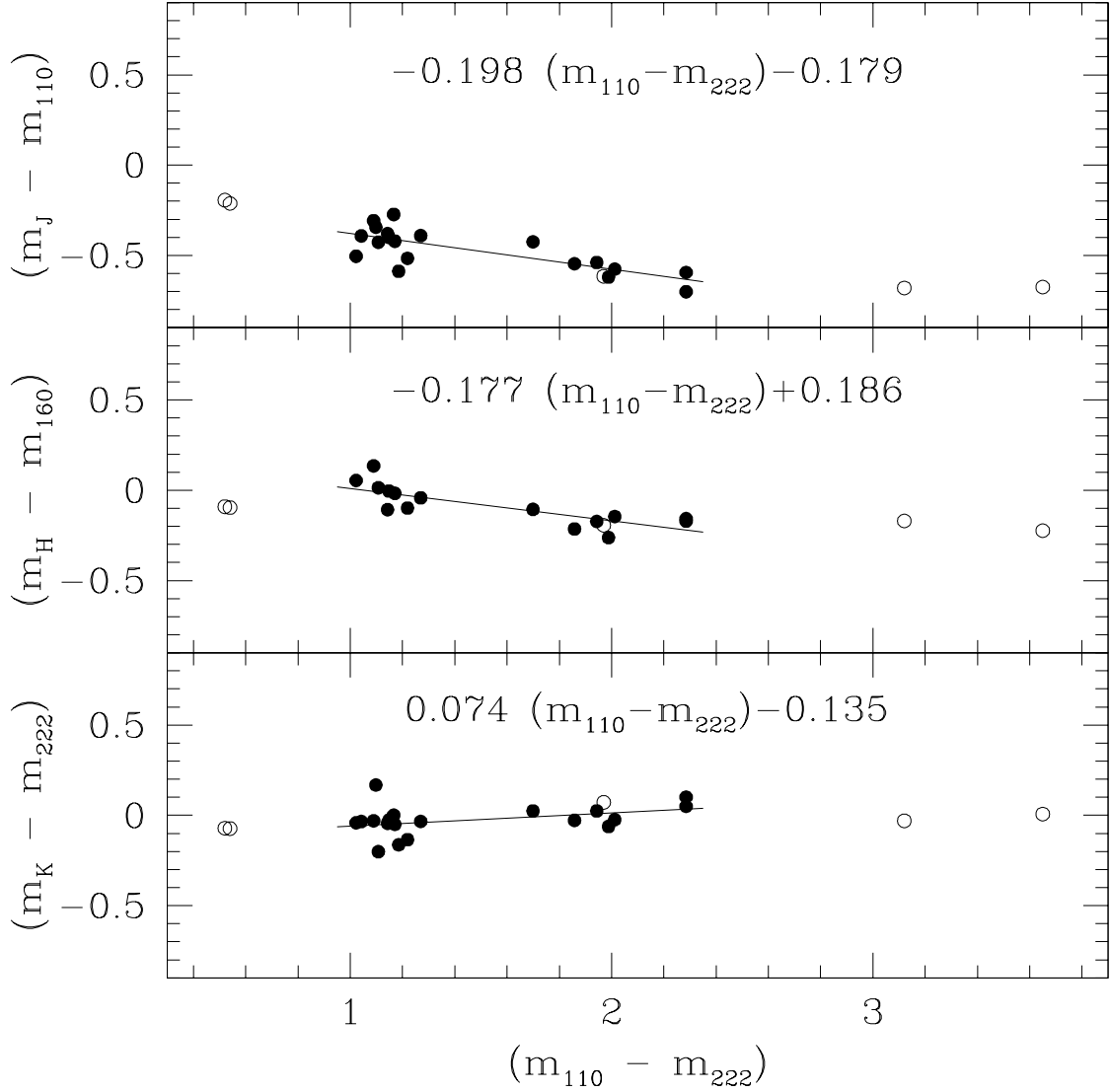}
\figcaption{Difference between NICMOS PHOTFNU \& ZP(VEGA) calibrated measurements and ground-based observations on the CIT/CTIO photometric system.  Filled circles represent our data (Table 3), and open circles the STScI NICMOS standards (Table 4).  The lines are linear fits to our data, giving half weight to points with $m_K > 10$, which in this plot also corresponds to $(m_{110} - m_{222}) < 1.6$. 
\label{Fig_02}} 
\end{figure}

\placefigure{Fig_02}

Using the PHOTFNU \& ZP(VEGA) keywords to calculate $m_{nicmos}$:
\begin{eqnarray} \begin{array}{cc}
m_J = m_{110} - (0.198 \pm 0.036) (m_{110} - m_{222}) - (0.179 \pm 0.060) &			 		 \\
m_H = m_{160} - (0.177 \pm 0.037) (m_{110} - m_{222}) + (0.186 \pm 0.065) & \;\;\;\; 1.0 < (m_{110}-m_{222}) < 2.3 \\
m_K = m_{222} + (0.074 \pm 0.037) (m_{110} - m_{222}) - (0.135 \pm 0.061) &					  \\
\end{array} \end{eqnarray} 

\begin{eqnarray} \begin{array}{cc} 
m_J = m_{110} - (0.344 \pm 0.063) (m_{110} - m_{160}) - (0.091 \pm 0.077) & \;\;\;\; 0.8 < (m_{110}-m_{160}) < 1.6 \\
m_H = m_{160} - (0.305 \pm 0.065) (m_{110} - m_{160}) + (0.259 \pm 0.081) &				 	 \\
\end{array} \end{eqnarray} 

Using the PHOTFLAM \& {\em our} PHOTZPT (from Table 2) to calculate $m_{nicmos}$:
\begin{eqnarray} \begin{array}{cc} 
m_J = m_{110} - (0.198 \pm 0.036) (m_{110} - m_{222}) \pm 0.058 &					 	\\
m_H = m_{160} - (0.177 \pm 0.037) (m_{110} - m_{222}) \pm 0.063 & \;\;\;\; 0.9 < (m_{110}-m_{222}) < 2.2 	\\
m_K = m_{222} + (0.074 \pm 0.037) (m_{110} - m_{222}) \pm 0.059 & 					 	\\
\end{array} \end{eqnarray} 

\begin{eqnarray} \begin{array}{cc} 
m_J = m_{110} - (0.344 \pm 0.063) (m_{110} - m_{160}) - (0.026 \pm 0.054) & \;\;\;\; 0.4 < (m_{110}-m_{160}) < 1.2 	\\
m_H = m_{160} - (0.305 \pm 0.065) (m_{110} - m_{160}) - (0.027 \pm 0.058) &					 	\\
\end{array} \end{eqnarray} 

Going directly from RAW instrumental nicmos magnitudes:
\begin{eqnarray} \begin{array}{cc} 
m_J = m_{110} - (0.198 \pm 0.036) (m_{110} - m_{222}) + (21.754 \pm 0.030) & 						\\
m_H = m_{160} - (0.177 \pm 0.037) (m_{110} - m_{222}) + (21.450 \pm 0.028) & \;\;\;\; -1.3 < (m_{110}-m_{222}) < 0.0 	\\
m_K = m_{222} + (0.074 \pm 0.037) (m_{110} - m_{222}) + (20.115 \pm 0.031) &  						\\
\end{array} \end{eqnarray} 

\begin{eqnarray} \begin{array}{cc} 
m_J = m_{110} - (0.344 \pm 0.063) (m_{110} - m_{160}) + (22.054 \pm 0.034) & \;\;\;\; 0.1 < (m_{110}-m_{160}) < 0.9 	\\
m_H = m_{160} - (0.305 \pm 0.065) (m_{110} - m_{160}) + (21.715 \pm 0.037) &  						\\
\end{array} \end{eqnarray}

These transformations are required due to the differences between CIT/CTIO and NICMOS filter bandpasses.  The NICMOS F110W filter is more than twice as wide as the $J$ filter and extends 0.3 \micron\ bluer.  The F160W filter is about 35\% wider than the $H$ filter and extends 0.1 \micron\ bluer, while the F222M filter is about half as wide as the $K$ filter.  Since NICMOS observations are sensitive to wavelengths at which radiation is completely absorbed by the Earth's atmosphere, any transformation between NICMOS and ground-based systems will only be valid for stars with similar spectral features.  We therefore must emphasize that our transformations are only for late-type stars with molecular absorption bands, especially H$_2$O and CO, and would not accurately transform heavily reddened blue stars for example.

STScI has observed six stars with NICMOS in order to establish a transformation between the F110W, F160W, and F222M filters and their ground-based counterparts $J,H,K$.  Only five of these stars have accurate $J,H,K$ ground-based measurements available (Table 4); these are over-plotted in Figure 3.  The NICMOS measurements have been calibrated on the ZP(VEGA) system.  The one star which lies in our color range is in excellent agreement with our observations and the other four allow us to estimate an extension of our transformation beyond the color limits of our sample.  The one potential problem with considering the STScI stars outside our color range is that they are not all late-type stars with significant molecular absorption bands.  The two stars blue-ward of our sample are both G stars, and except for the $H$ band, are consistent with our derived transformation.  The two stars red-ward of our sample consist of a heavily reddened B star and an M3 star, which seem to indicate that a change in slope occurs around $(m_{110}-m_{222}) \sim 2.5$, at which point the transformation slopes drop to zero, maintaining a constant offset.  These offsets are approximately -0.68, -0.19, and -0.01 for $J$,$H$, and $K$ respectively.

Low-resolution infrared spectra of six late-type stars with colors in the range $0.6 < (J-K) < 1.8$ and $\alpha$ CMa (A1 V) were taken above the atmosphere with a balloon-borne telescope (\cite{Woo64}).  Convolving these spectra with the band-passes of the HST filters and instrument throughput, we simulate NICMOS photometric observations.  Simulated ground-based photometry was produced by convolving the same spectra with $JHK$ filter band-passes and a transmission profile of the Earth's atmosphere.  This analysis yields slopes similar to the ones we derived above using actual photometric observations (Figure 3), in that the F110W and F160W filters give magnitudes too faint for redder stars, while the F222M filter gives magnitudes too bright for redder stars.  In Figure 3 the late-type stars are represented by filled circles and are the only points used for the linear least-squares fit.  It is important to note that slopes obtained from the simulation are close to those of our transformations given in eqn. 4 (the intercepts are merely zero-point shifts and are irrelevant in this analysis).  The main uncertainty comes from the need to extrapolate the Woolf \etal\ (1960) spectra which only go down to $\sim 0.96 \micron$ to the lower transmission limit of the NICMOS F110W filter at $\sim 0.76 \micron$.

In order to simulate the heavily reddened B star in the STScI standards, we have artificially reddened the A star $\alpha$ CMa with $A_J$ of 3.5 and 4.0, using the NIR extinction law from 
Mathis (1990) of $A_{\lambda} = A_J ( \frac{\lambda}{1.25\mu m} )^{-1.70}$.  These are the two open circles which lie at $(m_{110}-m_{222})$ of $\sim$ 2.7 and 3.1.  This A star seems to confirm that the transformation is valid much bluer than our sample for earlier type stars, while the reddened A star also shows the same change in slope which was indicated by the STScI sample.  Thus the reddened A star validates our technique of simulating photometry in that we can reproduce what was observed not only for the late-type stars but for the STScI reddened B star as well.

\begin{figure}
\plotone{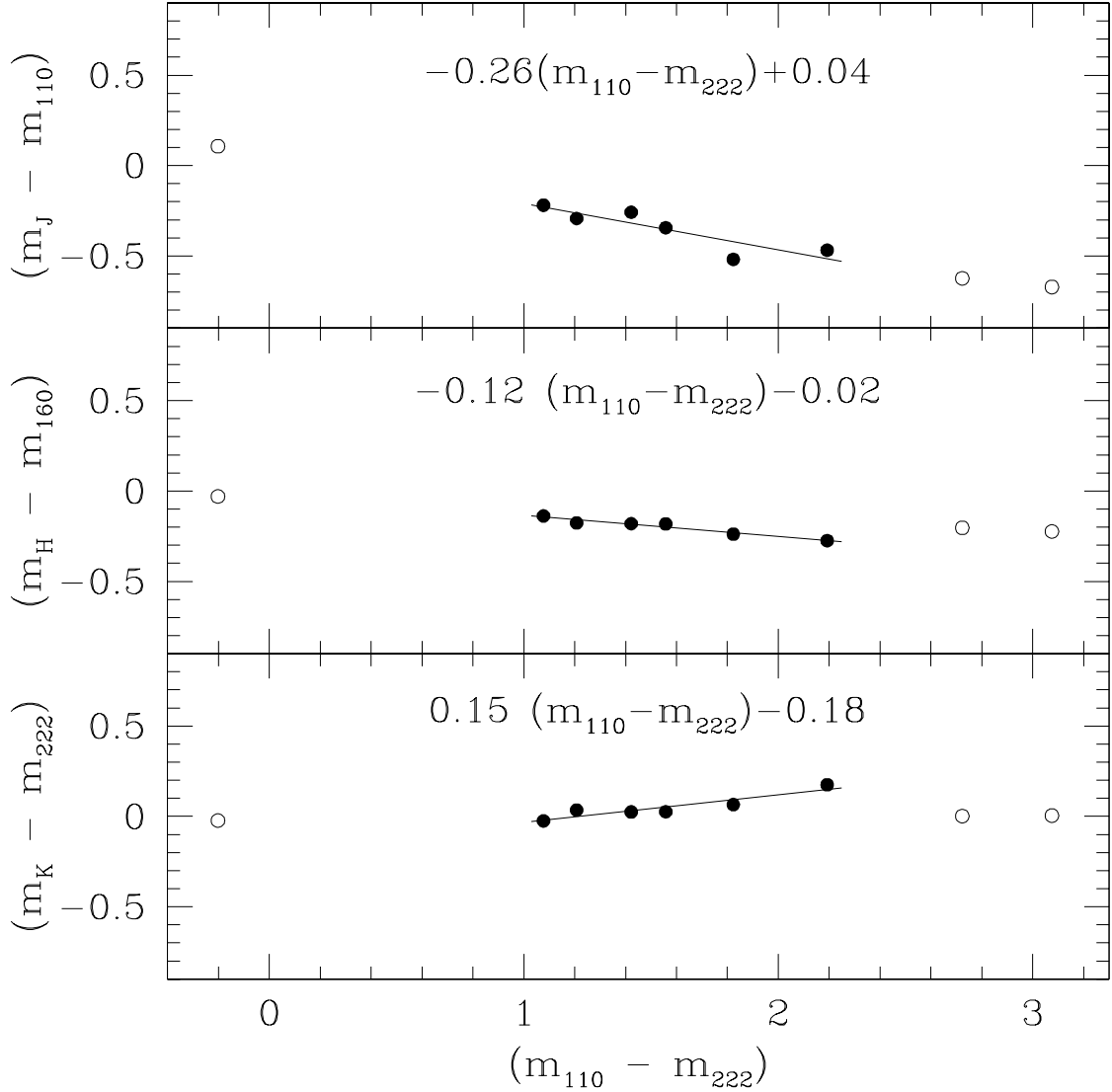}
\figcaption{Simulated photometry using low resolution IR spectra taken above the atmosphere (\cite{Woo64}) convolved with either ground-based $JHK$ filters and the atmospheric transmission or with NICMOS F110W,F160W,F222M filters (Table 5).  Filled dots are late-type stars, and are the only points used in the linear fit.  The open circles are an A star (left) and the same A star reddened with an $A_J$ of 4.0 and 4.5 (right).  Note the similarity of these slopes with the slopes we derive for our data (zero point shifts are arbitrary).
\label{Fig_03}}
\end{figure}

\placefigure{Fig_03}

\begin{deluxetable}{l r r r r | l r r r r r r r} 
\tablecaption{Observations}
\footnotesize
\tablewidth{0pt}
\tablehead{
	\colhead{}	&
	\colhead{}	&
	\multispan3{Ground-based} &
	\colhead{}	&
	\colhead{}	&
	\multispan5{Hubble Space Telescope \& NICMOS camera 2} \\
	\colhead{Field} &  
	\colhead{ID}	&
	\colhead{J}	&
	\colhead{H} 	& 
	\colhead{K} 	& 
	\colhead{Field} & 
	\colhead{ID} 	& 
	\colhead{F110W} & 
	\colhead{error}	&
	\colhead{F160W} & 
	\colhead{error}	&
	\colhead{F222M} &
	\colhead{error} 
}
\startdata
BW4b &  135 &  14.899 &  14.264 &  14.092 & BW1 &  25 & 15.278 & 0.029 & 14.372 & 0.027 & 14.136 & 0.050 \\
BW4b &  138 &  14.920 &  14.395 &  14.144 & BW1 &  43 & 15.319 & 0.030 & 14.400 & 0.027 & 14.171 & 0.051 \\
BW4b &  241 &  16.028 &   0.000 &  15.343 & BW1 &  64 & 16.420 & 0.050 & 15.595 & 0.047 & 15.378 & 0.089 \\
BW4b &  406 &  17.017 &  16.506 &  16.217 & BW1 &  66 & 17.438 & 0.080 & 16.523 & 0.072 & 16.267 & 0.134 \\
BW4b &  475 &  17.223 &   0.000 &  16.462 & BW1 &  72 & 17.810 & 0.098 & 16.810 & 0.082 & 16.625 & 0.158 \\
BW4b &  349 &  16.544 &  16.292 &  15.985 & BW1 & 135 & 17.048 & 0.066 & 16.238 & 0.063 & 16.026 & 0.120 \\
BW4b &  385 &  16.744 &   0.000 &  16.157 & BW1 & 162 & 17.087 & 0.068 & 16.179 & 0.061 & 15.989 & 0.117 \\
BW4b &  358 &  16.824 &  16.391 &  16.012 & BW1 & 187 & 17.131 & 0.069 & 16.256 & 0.064 & 16.042 & 0.121 \\
BW4b &  115 &  14.718 &  14.084 &  13.805 & BW1 & 197 & 15.108 & 0.027 & 14.126 & 0.024 & 13.839 & 0.043 \\
BW4b &  460 &  17.262 &  16.831 &  16.380 & BW1 & 226 & 17.689 & 0.090 & 16.817 & 0.083 & 16.581 & 0.155 \\
BW4b &  413 &  17.121 &   0.000 &  16.229 & BW1 & 276 & 17.394 & 0.078 & 16.440 & 0.069 & 16.228 & 0.131 \\
BW4b &  181 &  15.602 &  15.053 &  14.764 & BW1 & 284 & 16.126 & 0.043 & 15.161 & 0.038 & 14.902 & 0.071 \\
BMB  &  200 &  10.340 &   9.390 &   9.090 & BW2 &   6 & 10.765 & 0.007 &  9.497 & 0.005 &  9.066 & 0.010 \\
BMB  &  179 &   8.600 &   7.510 &   7.010 & BW3 &   3 &  9.194 & 0.003 &  7.668 & 0.002 &  6.909 & 0.003 \\
BMB  &  142 &   8.575 &   7.523 &   7.040 & BW4 &   6 &  9.276 & 0.003 &  7.695 & 0.002 &  6.991 & 0.003 \\
BMB  &  176 &   8.690 &   7.700 &   7.310 & BW5 &   7 &  9.229 & 0.003 &  7.873 & 0.002 &  7.286 & 0.003 \\
BMB  &  170 &  10.260 &   9.270 &   8.920 & BW6 &   9 & 10.806 & 0.007 &  9.485 & 0.004 &  8.948 & 0.007 \\
BMB  &  165 &   9.680 &   8.660 &   8.220 & BW7 &   7 & 10.256 & 0.005 &  8.806 & 0.003 &  8.244 & 0.005 \\
BMB  &  134 &   9.630 &   8.610 &   8.200 & BW8 &  10 & 10.250 & 0.005 &  8.873 & 0.003 &  8.262 & 0.006 \\
\enddata
\normalsize
\end{deluxetable}

\begin{deluxetable}{l c c c c c c c c c c c } 
\tablecaption{STScI NICMOS Standard Star Observations }
\footnotesize
\tablewidth{0pt}
\tablehead{
	\colhead{Star}	&
	\colhead{Type}	&
	\colhead{J}	&  
	\colhead{error}	&  
	\colhead{H}	&  
	\colhead{error}	&  
	\colhead{K}	&  
	\colhead{error}	&  
	\colhead{F110W}	&
	\colhead{F160W}	&
	\colhead{F222M}	&
	\colhead{error \tablenotemark{a}}	
}
\startdata
P330E    \tablenotemark{b} & G    & 11.816 & 0.007 & 11.479 & 0.005 & 11.419 & 0.007 & 12.01 & 11.57 & 11.49 & 0.02 \\
P177D    \tablenotemark{b} & G    & 12.258 & 0.012 & 11.924 & 0.008 & 11.857 & 0.013 & 12.47 & 12.02 & 11.93 & 0.02 \\
BRI 0021 \tablenotemark{b} & M9.5 & 11.835 & 0.008 & 11.086 & 0.007 & 10.552 & 0.010 & 12.45 & 11.28 & 10.48 & 0.02 \\
OPH-S1   \tablenotemark{c} & B    &  8.800 & 0.015 &  7.270 & 0.015 &  6.330 & 0.015 &  9.48 & 7.44  &  6.36 & 0.02 \\
CSKD-12  \tablenotemark{b} & M3   & 11.585 & 0.007 &  9.506 & 0.005 &  8.617 & 0.007 & 12.26 & 9.73  &  8.61 & 0.02 \\
\tablenotetext{a}{for F110W, F160W \& F222M}
\tablenotetext{b}{\cite{Per98}}
\tablenotetext{c}{\cite{Eli82}}
\enddata
\normalsize
\end{deluxetable}

\begin{deluxetable}{l c c } 
\tablecaption{\cite{Woo64} Balloon Observations}
\footnotesize
\tablewidth{0pt}
\tablehead{
	\colhead{Star}	&
	\colhead{Type}	&
	\colhead{$m_{110}-m_{222}$}
}
\startdata
$\alpha$ CMa   & A1 V       		& -0.202 \\
$\rho$ Per     & M4 II-III  		&  1.077 \\
$\alpha$ Ori   & M2-3 Iab   		&  1.208 \\
$\alpha$ Tau   & K5 III     		&  1.422 \\
$\mu$ Gem      & M3 III     		&  1.558 \\
R Leo          & M8         		&  1.823 \\
o Cet          & M9         		&  2.192 \\
Reddened CMa   & A1 V ($A_J=3.5$)  	&  2.723 \\
Reddened CMa   & A1 V ($A_J=4.0$)  	&  3.075 \\
\enddata
\normalsize
\end{deluxetable}

\section{Summary}

Our observations reveal that a color correction is required to transform HST NICMOS measurements to the CIT/CTIO photometric system.  The photometric keywords are listed in Table 2, and the transformations are given in section 4.  The validity of the transformation is confirmed with photometry simulated by convolution of filter transmission curves with late-type stellar spectra taken above the atmosphere.

The transformations we have derived are directly applicable ONLY to NIC2 because of differences in detectors, filters, and optics.  They may serve as a guide for observations with NIC1 and NIC3, but for precise photometry with these two cameras, a procedure similar to what we have done would need to be carried out.  These transformations are also only valid for late-type stars with molecular absorption bands because NICMOS is sensitive to wavelengths which never reach ground-based observatories.

\acknowledgments

We would again like to thank the TTRB for granting us time to reacquire the observations lost to the South Atlantic Anomaly.  Support for this work was provided by NASA through grant number GO-7826 from the Space Telescope Science Institute.  Daniela Calzetti, Antonella Nota, and Alfred Schultz at STScI provided help in understanding our data.  We would also like to thank Glenn Tiede for supplying us with the data for the BW4b field, Brian McLeod for his help in using NICRED, Marcia Rieke for her discussion on the STScI standards, and Paul Martini for helpful comments.

\clearpage

\clearpage

\clearpage

\end{document}